\documentstyle[aas2pp4,twoside,aabib]{article}
\input epsf

\newcommand{\oi}{O~{\sc i}}

\newcommand{\Jtco}{J=1$\rightarrow$0~$^{13}$CO}

\newcommand{\arcs}{{$^{\prime \prime}$}}
\newcommand{\arcm}{{$^{\prime}$}}
\newcommand{\kms}{km~s{$^{-1}$}}

\newcommand{\msol}{M{$_{\odot}$}}
\newcommand{\lsol}{L{$_{\odot}$}}

\begin{document}
\title{Supersonic Turbulence in the Perseus Molecular Cloud}

\author{Paolo Padoan}
\affil{Harvard University Department of Astronomy, Cambridge, MA 02138}

\author{John Bally \& Youssef Billawala}
\affil{Department of Astrophysics, Planetary, and Atmospheric Sciences,\\
       Center for Astrophysics and Space Astronomy,\\
       Campus Box 389, University of Colorado, Boulder CO 80309} 

\author{Mika Juvela}
\affil{Helsinki University Observatory, 
       T\"ahtitorninm\"aki, P.O.Box 14,
       SF-00014 University of Helsinki, Finland}

\author{\AA ke Nordlund}
\affil{Astronomical Observatory and Theoretical Astrophysics Center, \\
       Juliane Maries Vej 30, DK-2100 Copenhagen, Denmark}

\authoremail{ppadoan@cfa.harvard.edu}

\begin{abstract}

We compare the statistical properties of J=1$\rightarrow$0 $^{13}$CO
spectra observed in the Perseus Molecular Cloud with synthetic
J=1$\rightarrow$0 $^{13}$CO spectra, computed solving the non--LTE 
radiative transfer problem for a model cloud obtained as solutions 
of the three dimensional magneto--hydrodynamic (MHD)
equations. The model cloud is a randomly forced super--Alfv\'{e}nic 
and highly super--sonic turbulent isothermal flow.

The purpose of the present work is to test if idealized turbulent
flows, without self--gravity, stellar radiation, stellar outflows,
or any other effect of star formation, are inconsistent or not
with statistical properties of star forming molecular clouds.

We present several statistical results that demonstrate remarkable
similarity between real data and the synthetic cloud. Statistical
properties of molecular clouds like Perseus
are appropriately described by random supersonic and super--Alfv\'{e}nic
MHD flows. Although the description of gravity and stellar radiation are
essential to understand the formation of single protostars and the effects
of star formation in the cloud dynamics, the overall description of the
cloud and of the initial conditions for star formation can apparently be
provided on intermediate scales without accounting for gravity,
stellar radiation, and a detailed modeling of stellar outflows 

We also show that the relation between equivalent line width and 
integrated antenna temperature indicates the presence of a relatively
strong magnetic field in the core B1, in agreement with Zeeman
splitting measurements.

\end{abstract}

\keywords{
turbulence -- ISM: kinematics and dynamics -- magnetic fields --
individual (Perseus Cloud); radio astronomy: interstellar: lines
}

\section{Introduction}

The structure and dynamics of molecular clouds (MCs) has been the subject
of intensive investigation during the past two decades. Since MCs are the
sites of star formation, their properties provide insights into the
initial conditions needed to form stars. On the other hand, energy
injected into the clouds by young stars is also believed to play a crucial
role in the cloud dynamics and evolution (e.g. Whitworth 1979; Bally \&
Lada 1983; Miesch \& Bally 1994). The observed state of a cloud represents
the balance between energy injection and dissipation. 

Observations show that cloud internal motions are highly supersonic,
incoherent, and random (e.g. Bally et al.\ 1987, 1989, 1990).
Furthermore, the cloud structure is complex and by many measures is
suggestive of the presence of random flows. Indeed, many authors have
attempted to describe clouds using various models of turbulence (e.g.
Ferrini, Marchesoni \& Vulpiani 1983; Henriksen \& Turner 1984; Scalo
1987; Fleck 1988; Fleck 1996) inspired by the scaling laws observed in the
interstellar medium (Larson 1981). More recently, there have been several
attempts to describe cloud morphology and velocity fields using fractals
(e.g. Scalo 1990; Falgarone \& Phillips 1991; Falgarone, Phillips \&
Walker 1991; Falgarone 1992; Larson 1995; Elmegreen 1997). 

Most theoreticians have argued that clouds are supported by magnetic
fields (Mestel 1965; Strittmatter 1966; Parker 1973; Mouschovias 1976a,
b; McKee \& Zweibel 1995) and interpreted the observed line--widths in terms
of magneto--hydrodynamic (MHD) waves (Arons \& Max 1975; Zweibel \&
Josafatsson 1983; Elmegreen 1985, Falgarone \& Puget 1986). However,
comparison of the extinction statistics with recent models of supersonic
and super--Alfv\'{e}nic random flows suggests that such models may provide
a superior description of MCs (Padoan, Jones \& Nordlund 1997).
Zeeman splitting measurements (for example Crutcher et al.\ 1993; 
Crutcher et al.\ 1996) have also
been shown to be consistent with the predictions of a super--Alfv\'{e}nic
random flow (Padoan \& Nordlund 1998). 

In this paper, we compare some observed properties of the Perseus
molecular cloud with the predictions of super--sonic and super--Alfv\'{e}nic
random flows. We start with a self--consistent MHD simulation of a random
flow in a three dimensional $128^3$ grid. We solve the
radiative transfer problem using a non--LTE Monte Carlo approach (Juvela
1997) to compute $^{13}$CO synthetic spectra on a 90 by 90 cell grid with
60 velocity channels. The experiment we use for the present paper is
the same as model Ad1 in Padoan \& Nordlund (1998), where a detailed 
description of the code is given. The method of obtaining synthetic spectra 
from our numerical simulations of MHD turbulence is presented in 
Padoan et al.\ (1998).

Super--sonic random flows can be shown to offer a natural
interpretation of statistical and morphological properties of molecular
clouds, at least for clouds that are not very active regions
of star formation. For example, Padoan et al.\ (1998) calculate 
values of CO line intensity ratios and line width ratios, using 
synthetic spectra, that are very close to the values measured 
by Falgarone et al.\ (1991) and by Falgarone \& Phillips (1996) 
in quiescent regions. It is normally believed that the study of 
molecular clouds that are actively forming stars requires instead 
a complicated modeling of the effect of star formation, such as
energy injection from molecular outflows, HII regions, and so on.
If this were true, it would be very difficult to learn general
statistical properties of star forming molecular clouds, directly 
from numerical simulations of MHD flows. The present paper 
addresses this problem, by comparing statistical properties
of super--sonic and super--Alfv\'{e}nic turbulent flows with the
same properties in the Perseus molecular cloud complex, which is
a well known star formation site. 
The comparison between the model and the observations
shows that a reasonable description of the kinematics and structure
of molecular clouds can be provided by a turbulent flow model,
that does not account for gravity, stellar radiation
fields, stellar outflows, or any other effect of star formation.

\section{The Perseus Molecular Cloud}

Observations of the Perseus molecular cloud complex were made from 1985 to
1991 with AT\&T Bell Laboratories 7~m offset Cassegrain antenna in
Holmdel, New Jersey. Approximately a 6$^\circ$ by 4$^\circ$ region was
surveyed in the 110 GHz line of \Jtco . The observations were made on a
1\arcm\ grid with a 100\arcs\ beam yielding 33,000 spectra, each with
128~$\times$~100~kHz channels (0.273~\kms ). See Billawala et al.\ (1997)
for further details of observations and discussion of the cloud structure. 

The Perseus molecular cloud is roughly a 6$^\circ$ by 2$^\circ$ region
located at a distance estimated to range from 200 to 350~pc (Herbig \&
Jones 1983; C\^ernis 1990) and has a projected size of 30 by 10~pc (see
figure 1). The cloud is part of the larger Taurus--Auriga--Perseus complex
and lies below the galactic plane ({\it b} $\sim\ -20^\circ $) with an
estimated mass of 10$^4$~\msol\ (Billawala et al.\ 1997). 

The complex has produced some high mass stars at its eastern end which lie
near the centroid of the Perseus OB2 association, a loose grouping of 5 to 10
O and B stars, that is about 7 million years old (Blaauw 1991). Perseus
OB2 has blown a 20$^\circ$ diameter (100~pc) supershell into the
surrounding interstellar medium that can be seen in 21~cm H~${\sc i}$ data
(Heiles 1984). The shell is partially superimposed on the Perseus
molecular cloud and may be interacting with it. Perseus contains two young
stellar clusters and a background of relatively inactive molecular gas
that has formed few stars. 

Near the eastern end is the young ($<$ 7~Myr old) cluster IC348, which has
several hundred members and may still be forming (Strom, Grasdalen \&
Strom 1974; Lada, Strom \& Myers 1993; Lada \& Lada 1995). Omicron Persei,
a B1 III star, located near the center, is believed to be a few parsecs 
in front of the cloud
(Bachiller et al.\ 1987). Lada \& Lada (1995) estimate a cluster population
of $\approx $ 380 stars with a highly centralized distribution falling off
inversely with distance from the inner 0.1~pc diameter core which contains
about $\sim$~200~\msol\ of stars and a density of 220~\msol~pc$^{-3}$.
Star formation has been going on for at least 5 to 7~Myr at a roughly
constant rate and with an efficiency of $\sim~50\%$. 

A second and even younger cluster is embedded in the Perseus cloud several
degrees to the west near the reflection nebula NGC1333. This region
contains the most active site of ongoing star formation in Perseus.
NGC1333 contains an embedded $<$~1~Myr old infrared cluster, a large
number of molecular outflows (Sandell \& Knee 1998), Herbig--Haro objects
(Bally, Devine \& Reipurth 1996), and shock excited near--infrared H$_2$
emission regions (Aspin, Sandell \& Russell 1994; Hodapp \& Ladd 1995; 
Sandell et al.\ 1994). Lada, Alves \& Lada (1996) find 143 young stars 
in a 432 square
arc--minute region with most being members of two prominent sub--clusters. 
More than half of these stars display infrared excess emission, a
signature of young stellar objects. Lada et al.\ estimate the age of the
cluster to be less than $10^6$ years with a star formation rate of
$4~\times~10^{-5}$~\msol~yr$^{-1}$. 

Away from IC348 and NGC1333, the Perseus cloud contains about a dozen
dense cloud cores with low levels of star formation activity including the
dark cloud B5 (Langer 1989; Bally, Devine, \&  Alten  1996) at its eastern
end, the cold core B1 for which a magnetic field has been measured by
means of OH Zeeman (Goodman et al.\ 1989; Crutcher et al.\ 1993), and the
dark clouds L1448 and L1455 on the west end (Bally et al.\ 1997). 

The cloud exhibits a wealth of sub--structures such as cores, shells, and
filaments and dynamical structures including stellar outflows and jets,
and a large scale velocity gradient. While emission at the western end of
the complex lies mostly near v$_{lsr}$~=~4~\kms\ (L1448), the eastern end
of the complex has a velocity of v$_{lsr}$~=~10.5~\kms (B5). This may be
an indication that we are seeing several smaller clouds partially
superimposed along the line of sight (C\^ernis 1990). 

\section{The Model}

A synthetic molecular cloud map containing grids
of 90~$\times$~90 spectra of \Jtco\  has been computed with a
non--LTE Monte Carlo code (Juvela 1997), starting from density and velocity
fields that provide realistic descriptions of the observed physical
conditions in MCs. The density and velocity fields are obtained as
solutions of the magneto--hydrodynamic (MHD) equations in a 128$^3$ grid
and in both super--Alfv\'{e}nic and highly super--sonic regimes of a random
isothermal flow. The numerical experiment is the same as model Ad1 in
Padoan \& Nordlund (1998), where details about the code and different 
experiments are given. The method of calculating synthetic spectra from 
numerical simulations of MHD turbulence is presented in Padoan et al.\ (1998).

The numerical MHD flow that is used in the present work (model Ad1 of
Padoan \& Nordlund 1998) has initial rms Mach number ${\cal M}=10.6$ 
and initial rms alfv\'{e}nic Mach number ${\cal M}_A=10.6$, where 
${\cal M}$ is the ratio of the rms velocity of the flow divided by the 
speed of sound, and ${\cal M}_A$ is the rms velocity of the flow divided 
by the Alfv\'{e}n velocity. The flow is randomly driven, but the driving 
and the initial conditions are such that the rms velocity decreases
for the first two dynamical times. The snapshot we use for the present work
corresponds to a time equal to almost two initial dynamical times, where
the Mach numbers are ${\cal M}=7.7$ and ${\cal M}_A=2.5$. The alfv\'{e}nic 
Mach number has decreased also due to the amplification of the magnetic energy
caused by the strong compressions and by the stretching of magnetic field lines,
in the initial evolution of the flow. Molecular clouds have internal random 
motions of similar Mach number on the scale of about 3.7~pc (assuming
a kinetic temperature of about 10~K), where the mean density is 
$<n>\approx540$~cm$^{-3}$. These are therefore the values of the linear size
and gas density that are used in the radiative transfer calculations.
The corresponding mass of the model is 1.7$\times$10$^3$~\msol\
(using a mean molecular weight of $\mu$~=2.6 to correct for He).

The gas density spans the range of values 
6.8~cm$^{-3}$--3.3$\times$10$^4$~cm$^{-3}$, which produces
column density values over a range of almost two orders of magnitude. 
Models similar to this have already been shown to reproduce observed 
statistical properties of molecular clouds (Padoan, Jones \& Nordlund 1997; 
Padoan et al.\ 1998; Padoan \& Nordlund 1998). Although
a version of our MHD code can include self--gravity, and the Monte Carlo
radiative transfer code can include the stellar radiation field,
we have chosen not to include gravity or any effect of star formation.
In fact, the purpose of the present work is to investigate if such a highly 
idealized description of the dynamics of molecular clouds is
consistent or not with statistical properties of star forming 
molecular clouds. While it is clear that gravity is eventually important
for the collapse of single protostars, and molecular outflows are important 
energy sources, it might be that the dynamics and statistical
properties of star forming molecular clouds are described rather well
by a randomly forced MHD super--sonic flow, without gravity and without
a specific energy injection mimicking stellar outflows.

\section{Results}

In order to compare the observed spectra with the synthetic ones,
it is necessary to eliminate the contribution of the noise to the 
value of the statistical quantities computed with the observed spectra.
This is especially true for the value of the kurtosis of the velocity 
profiles, since it is the statistics of highest order among the ones
we compute. The noise is treated in the following way. First, the noise
is estimated using velocity channels where no emission is apparently
detected. The average antenna temperature in such velocity channels is
-0.0009 K, and the rms value is 0.1556 K. Then the antenna temperature
is set equal to zero in all velocity channels where it is lower than 
the $2\sigma$ value of the noise, that is 0.3 K.
Finally, to avoid unphysical sharp features in the velocity profiles,
due to the antenna temperature clipping, the spectra are smoothed,
only in velocity channels with antenna temperature lower than 0.8 K, 
with a gaussian filter with a width of 5 velocity channels. We have 
verified that the final smoothing affects the statistical distributions 
very little, while the antenna temperature clipping is necessary to
obtain any agreement between observational data and synthetic spectra.
Although the choice of the clipping at the $2\sigma$ noise level
is somewhat arbitrary, we point out that this choice not only allows
an excellent fit of the distribution of kurtosis, but also allow
a match, at the same time, of all other statistical distributions
we have computed. A different clipping level, eliminates
the good fit of all statistics at the same time. 
It remains to be checked if the same 
clipping level is valid for observed spectra with any value of signal to
noise, or if spectra of different quality need to be treated with
a different clipping level.

The synthetic integrated antenna temperature maps, computed from 
the model clouds solving the radiative transfer problem along the
x, y and z axis, look rather filamentary (left panel of Fig.~\ref{fig2}).
The Perseus map (Fig.~\ref{fig1}), instead, does not show a very filamentary 
morphology. However, the resolution of the observations is inferior
to the resolution of the synthetic maps. To allow for a comparison, it is
necessary to degrade the resolution of the synthetic map to a level
comparable to the resolution in the observations. This has been done
by filtering the maps with a gaussian filter with width approximately equal
to twice the nominal size of the 100" observational beam. 
The morphology of the synthetic maps after filtering is not very different
from the observed morphology; the underlying filamentation is hardly visible.
In the same way, an intrinsic filamentation in the Perseus molecular cloud 
complex would be largely hidden from view in the present data, for lack of
resolution.

Examples of grids of spectra are shown in Fig.~\ref{fig3} and Fig.~\ref{fig4}.
The original synthetic maps contain 90$\times$90 spectra, but only a subset of
30$\times$30 spectra is plotted here. Model and real clouds both show
some spectra that are close to gaussian, and some other with multiple components.
These spectra can be used to compute several statistical properties of 
the clouds. In Fig.s~\ref{fig5}-\ref{fig7}, different statistics are shown for 
regions around L1448, NGC1333, and B1. In each figure the histograms 
computed with the synthetic spectra are also overplotted for comparison
(thick lines), and a small contour map of the selected region is shown.
Since the model is not computed with the purpose of fitting any 
specific cloud, the histogram of velocity centroids, $<v>$, is not expected to
match the observations. Such histograms depend in fact on the specific large
scale motions inside each cloud. 

The distribution of the standard deviation of single spectra, $\sigma(v)$,
(the dispersion of radial velocity in each line of sight), can also be a 
tricky quantity to fit with a theoretical model. 
It may happen that two or more
physically disconnected clouds are superposed on the line of sight and
produce a spectrum with multiple components, at different velocities,
which may cause a significant growth of the standard deviation. Multiple
components are found also in the synthetic spectra, but are due to the
superposition of structures that belong to the same cloud. The enhancement
of $\sigma(v)$ can be important when large regions are sampled, because the 
chance of random superposition of disconnected clouds is large.

All other histograms plotted in Fig.s~\ref{fig5}-\ref{fig7} do instead measure 
to what extent the model is consistent with the observations. Such statistics are
the probability distribution of skewness and kurtosis of the spectra, 
that constrains the shape of the spectra; the probability distribution of 
equivalent line width, $\Delta v$; the probability distribution of velocity 
integrated antenna temperature, $\int Tdv$ (roughly proportional to column 
density); the relation between equivalent line width and integrated 
temperature. The distributions are computed by measuring all the above 
quantities in each line of sight.

\subsection{L1448}

A region around the cloud core L1448 has been selected, that contains
also a part of the core L1455 (visible in the bottom left corner on 
the contour map in Fig.~\ref{fig5}).
The dark core L1448 contains a number of young stars and  Herbig--Haro
objects, but it is a relatively quiescent region in the cloud complex.
Fig.~\ref{fig5} shows that the statistics of the theoretical model match 
almost exactly the statistics of the spectra of L1448.

In Fig.~\ref{fig11}, the same statistical properties are plotted
against each other, and again the agreement between the model and the 
observations is excellent. For example, both observed and synthetic
spectra with large kurtosis (larger than gaussian) tend to have extreme 
values of skewness (negative or positive). Only the relations between 
kurtosis and line width seem to be significantly different
between theory and observations. Similar results are obtained for other
regions in the Perseus complex, and the equivalent of Fig.~\ref{fig11}
for other regions is not shown in the following. 

Figures \ref{fig5} and \ref{fig11} show that statistical properties
of spectra observed around the core L1448 are well reproduced by the 
present idealized model, and therefore are likely to be related to 
super--sonic and super--Alfv\'{e}nic turbulence present in the
molecular gas.

\subsection{NGC1333}

The region around 
the reflection nebula NGC1333 is instead the most active site of ongoing
star formation in the Perseus cloud complex. Molecular outflows, Herbig--Haro
objects, and shock excited H$_2$ emission regions have been observed.
The comparison of the model with NGC1333, in  Fig.~\ref{fig6}, shows that 
NGC1333 has slightly larger line width than the model. This is not surprising,
since the model is not tailored to any particular cloud core. Consistently
with the larger line width, the distribution of integrated temperature
is also a bit broader than in the model. In fact, it is a feature of our
models that flows with larger velocity dispersion (hence larger $\Delta v$)
produce broader distributions of integrated temperature (Padoan et al.\ 1998),
than flows with smaller velocity dispersion. The histograms of skewness
and kurtosis match rather well the model, and the relation between 
equivalent line width and integrated antenna temperature is virtually
identical to the theoretical one. 

Even the very active region around NGC1333 has therefore statistical 
properties that are well described by the present idealized model
of the dynamics of molecular clouds.

\subsection{B1}

The core B1 is very interesting because a rather strong magnetic field 
strength has been detected there (Goodman et al.\ 1989; Crutcher et al.\ 1993). 
The match between model and 
observations is again very good (Fig.~\ref{fig7}), apart from the distribution
of integrated antenna temperature and for the relation between line width and
integrated temperature. The distribution of integrated temperature is
perhaps affected by the very small size of the region selected around the core,
that does not include a significant area of very low column density. Even
in the synthetic maps one can select a small region around a core or a
filament and obtain a distribution of integrated temperature
similar to the one observed in the core B1.

The line width in B1 does not grow with integrated temperature, above 
the value of 2~K~km/s, and even decreases slightly up to the value
of 8~K~km/s. This behavior of the line width is observed in our models
with relatively strong magnetic field (rough equipartition of kinetic
and magnetic energy), as shown in Padoan \& Nordlund (1998).
In the same work it is also shown that the formation of some cores with 
relatively strong magnetic field is also predicted in the context of
a super--Alfv\'{e}nic model for the cloud dynamics, such as the present model.
The relation between equivalent line width and integrated antenna
temperature in the core B1 is therefore in agreement with the strong
field detected in this core with Zeeman splitting measurements,
and is consistent with our super--sonic and super--Alfv\'{e}nic turbulence 
model for the dynamics of molecular clouds.

\section{Discussion and Results}

\subsection{Observational and Synthetic Molecular Line Spectra}

Several authors have developed theoretical models of dark clouds
to interpret molecular line observations. Although the earliest
interpretation of the carbon monoxide line profiles was that 
molecular clouds are collapsing (Goldreich \& Kwan 1974; Liszt et al.\ 
1974), it was soon realized that systematic motions different than collapse,
or random motions were more likely explanations for the line profiles
(Zuckerman \& Evans 1974; Leung \& Liszt 1976; Baker 1976; Kwan 1978; 
Albrecht \& Kegel 1987). Models have then been improved with the use
of clumpy density distributions, where the volume filling fraction of 
the clumps and their internal density are the main parameters
(Kwan \& Sanders 1986; Tauber \& Goldsmith 1990; Tauber, Goldsmith 
\& Dickman 1991; Wolfire, Hollenbach \& Tielens 1993; Robert \& Pagani 
1993; Park \& Hong 1995; Park, Hong \& Minh 1996) and with the
use of fractal density distributions (Juvela 1997).
    
\nocite{Goldreich+Kwan74,Liszt+74,Kwan78}
\nocite{Zuckerman+Evans74}
\nocite{Leung+Liszt76,Baker76,Dickman78,Kwan+Sanders86}
\nocite{Albrecht+Kegel87,Tauber+Goldsmith90,Tauber+91}
\nocite{Wolfire+93,Robert+Pagani93,Park+Hong95,Park+96,Juvela97}

Although velocity and density distributions used in such models 
are not related in any way to solutions of the fluid--dynamic
equations, it was generally believed that some kind of MHD waves
or hydro--dynamic turbulence should be at the origin of the 
random motions. Some studies have attempted to relate 
molecular cloud motions, as probed by line profiles, to turbulence. 
Falgarone \& Phillips (1990) studied molecular line profiles
\nocite{Falgarone+Phillips90} of different sources, and found excess 
emission in the line wings, relative to a Gaussian distribution 
(see also Blitz, Magnani \& Wandel 1988).
\nocite{Blitz+88}
Miesch \& Bally (1994) computed autocorrelation and structure
functions of line centroid velocities in five nearby mlecular clouds,
and compare their results with phenomenological theories of turbulence.
\nocite{Miesch+Bally94}
Miesch \& Scalo (1995) looked at the histograms of emission line centroids,
\nocite{Miesch+Scalo95}
and found nearly exponential tails in many cases. Lis et al.\ (1996) 
\nocite{Lis+96}
studied the statistics of line centroids and of centroid increments, 
and found non-Gaussian behavior especially in the histograms of the 
velocity centroid increments.

Stenholm \& Pudritz (1993) and \nocite{Stenholm+Pudritz93}
Falgarone et al.\ (1994) calculated synthetic molecular spectra from fluid 
\nocite{Falgarone+94}
models of clouds.  Stenholm \& Pudritz (1993) used a sticky particles code 
with an imposed spectrum of Alfv\'{e}n waves (Carlberg \& Pudritz 1990). 
\nocite{Carlberg+Pudritz90}
They calculated line profiles under the LTE assumption. 
Falgarone et al.\ (1994) did not solve the radiative transfer problem but
calculated density weighted radial velocity profiles, using velocity and 
density fields from a numerical simulation of compressible turbulence 
by Porter, Pouquet \& Woodward (1994).
\nocite{Porter+94} 
Padoan et al.\ (1998) calculated several maps of 90 by 90 synthetic spectra
of different transitions of various molecules, using models and 
methods similar to the ones used here.  Both the numerical models 
and the radiative transfer calculations constitute a significant 
improvement on the Falgarone et al.\ (1994) investigation.  The 
numerical model used in that work (Porter et al.\ 1994) had Mach 
numbers close to unity, and therefore a relatively small density 
contrast, inconsistent with the density
structure of molecular clouds, and in contradiction with the many models of line 
profiles that have proved the importance of the clumpy structure of molecular 
clouds (see references above). Moreover, while in Falgarone et al.\ (1994) the
line profiles are calculated simply as density weighted radial velocity profiles,
in Padoan et al.\ (1994) the line profiles are calculated by solving the
radiative transfer problem, and line intensity ratios and line width ratios
of different molecules or of different transitions of the same molecules
are calculated, and shown to compare well with the observations. 

According to Falgarone et al.\ (1994) the non--Gaussian shape of
the line profiles in molecular clouds is the results of intermittency, and
shocks make a minor contribution to the structure of the velocity field.
Although this might be true for the simulations by Porter, Pouquet \& 
Woodward (1994), that are transonic and so are characterized more by 
vortex tubes than by shocks, it cannot be true in the case of the highly
super--sonic motions in molecular clouds. In the simulations of
highly super--sonic turbulence by Padoan et al.\ (1998), the flows
are characterized by a complex system of interacting shocks, that are 
responsible for the generation of large density contrasts, as observed
in molecular clouds. Several of these velocity discontinuities are present
along each line of sight through the cloud model, and must be an essential
ingredient in the integrated velocity profile. The correlation
between density and velocity fields is also essential, and can be 
studied appropriately only with highly super--sonic flows. 
The statement that intermittency is at the origin of the non--Gaussian
shapes has been critisized in Dubinski, Narayan \& Phillips (1995), 
\nocite{Dubinski+95} where it
is shown that non--Gaussian shapes arise trivially from steep power spectra,
as the Kolmogorov power spectrum, similar to the one obtained in the
simulation by Porter, Pouquet \& Woodward (1994). 
Finally, Falgarone et al.\ (1994) do not discuss the role played by the magnetic
field in the velocity profiles, since the magnetic field is not 
included in the calculations by Porter, Pouquet \& Woodward (1994).

The present work follows the same method as in Padoan et al.\ (1998),
but uses new runs with larger rms Mach number (up to 28), presented 
in Padoan \& Nordlund (1998). The rms Mach number of the model used in this
work, ${\cal M}=7.7$, is typical of motions inside molecular clouds, and
therefore there is no need to rescale the density
field in order to mimick a larger Mach number, as in Padoan et al.\ (1998).
It has been shown that the distributions of skewness and curtosis of the 
synthetic line profiles of single line of sight matches very well 
the same distributions for the line profiles observed in molecular
clouds of the Perseus complex. The match is very good also between
theoretical and observed integrated temperature distributions, and 
integrated temperature versus equivalent width relations. 
The Perseus complex has been chosen because it is a well known
site of ongoing star formation, and it is an important question
wether the effect of star formation activity change drastically
the statistical properties of molecular cloud motions, or if such motions
can still be described statistically with idealized models of turbulent flows,
without self gravity, radiation field, and stellar outflows. 

Padoan \& Nordlund (1998) proposed to use the relation between the 
integrated temperature and the equivalent width to estimate the dynamic
importance of the magnetic field in molecular clouds, following an earlier 
suggestion by Heyer, Carpenter \& Ladd (1996).
\nocite{Heyer+96} 
This relation is normally found to be consistent with the motions being
super--Alfv\'{e}nic (Heyer, Carpenter \& Ladd 1996; Padoan \& Nordlund 1998).
We have verified this also in the case of the Perseus complex: the equivalent
line width always grows with increasing integrated antenna temperature. Only
in the case of the core B1 have we found that the line width does not 
grow, or even decreases, with increasing integrated antenna temperature.
According to our models, this indicates the presence
of a strong magnetic field, with a rough equipartition of magnetic and
kinetic energy. The core B1 is in fact the single spot in the Perseus complex
where a relatively strong magnetic field has been succesfully detected
(Goodman et al.\ 1989; Crutcher et al.\ 1993), and is believed to be important in
the dynamics of the core. This example illustrates that the inclusion of the
magnetic field in a cloud model can be important to the interpretation of the 
line profiles, and gives further support to the idea that the dynamics of
molecular clouds is essentially super--Alfv\'{e}nic (Padoan \& Nordlund 1998), 
because if that was not the case, the line width versus integrated temperature relation
would always look like the one in the core B1, contrary to the observational
evidence, and because the formation of cores with strong magnetic field is
predicted in the super--Alfv\'{e}nic model (Padoan \& Nordlund 1998).

\subsection{Super--Sonic Turbulence in Molecular Clouds}

The model used in this work is consistent with general observed 
properties of dark clouds such as the existence of random and 
supersonic motions, but it is not tailored to the specific characteristics 
of the Perseus molecular cloud or individual cores. Only the general 
interstellar medium scaling laws (Larson 1981) have been used to fix 
the physical size of the cloud model, and the comparison
of the model with the observational data is justified by their
similar size, inner scale (resolution) and line widths.

The dynamic range of length scales covered by the MHD calculations extends
from 0.03~pc to 3.7~pc (128$^3$ grid--points). The radiative transfer 
calculations and the resulting spectra cover a linear scale range from 
0.04~pc to 3.7~pc (90$^3$ grid--points). However, numerical dissipation 
effectively degrades this resolution by a factor of two to an inner scale 
of about 0.06~pc. Thus the inner scale of the theoretical models is about
half of the linear resolution of the 100\arcs\ beam used in our observations. 
The Bell Labs 7~meter telescope beam has a linear resolution of about 0.14 pc 
(assuming a distance of 300~pc for the Perseus cloud) and the maximum extent 
of the Perseus cloud that has been mapped is about 30~pc. If the Perseus
cloud is interpreted as an elongated single cloud, its typical
thickness is about 4~pc; if it is interpreted as a line--of--sight
superposition of different clouds, then each cloud has a size of
about 4~pc. These sizes are also comparable to the linear size
of the numerical model. Moreover, average spectra from single 
components of the Perseus cloud, and line widths of single 
line--of--sights are also comparable to the ones in the model.

Giant molecular clouds (GMCs; M$~>~10^5$~\msol\ and L $\approx$ 20 to
50~pc) are gravitationally bound. This must be the case since the product
of the internal velocity dispersion squared times the density is one to
two orders of magnitude greater than the pressure of the surrounding
interstellar medium. Gravity must provide the restoring force on the
scale of a GMC, otherwise GMCs would be dispersed before they could form
any stars. Individual stars form from the gravitational contraction of
small cores with sizes of order 0.1~pc, while clusters of stars such as
IC348 and NGC1333 form from cores about 1~pc in diameter. Thus gravity
is certainly important on the very large scale of GMCs (L $>$ 20~pc) and
on the very small scales of dense star forming cores (L $\le$ 1~pc). Our
model and observations have linear dimensions intermediate between these
two scales. The results of this work show that on intermediate scales the
observed properties of molecular clouds and likely the initial conditions
for star formation can be appropriately described without explicitly
accounting for gravity. Self--gravitating cores may occasionally condense
from the random flow where a local over-density is produced statistically.
However, at any one time, only a fraction of the total mass is involved in
such condensations. If indeed self--gravitating cores condense out of
a complex density field generated by super--sonic turbulent
flows, then it might be possible to predict the
mass spectrum of the resulting condensations which produce stars. This may
be an important ingredient in constructing the initial stellar mass
function (Padoan, Nordlund \& Jones 1997). 

In this picture, the overall density structure inside a GMC is {\it not}
the result of gravitational fragmentation, but rather of the presence of
supersonic random motions together with the short cooling time of the
molecular gas. We predict that there are several shocks in the gas along
any line--of--sight with velocities comparable to the overall cloud
line--width. These shocks have velocities of order 1 to 10 \kms\ which are
not observable in the forbidden lines in the visible portion of the
spectrum. However, there are several far--infrared transitions such as the
157~$\mu$m fine--structure line of C$^+$, the 63~$\mu$m line of \oi \ , and
the 28, 17, and 12 $\mu$m v~=~0--0 lines of H$_2$ which are readily
excited by such shocks. The COBE and ISO satellites have shown that these
transitions together carry about 0.1\% of the total luminosity of a
typical galaxy (Wright et al.\ 1991; cf. Lord et al.\ 1996). 
We predict that
roughly 10\% of this emission is produced by the shocks discussed above
because these transitions are the important coolants in the post--shock gas. 

The dissipation of the kinetic energy of internal motions, $E_{GMC}$, by
shocks in a GMC produces a luminosity of order $L_{GMC} ~\approx ~ E_{GMC}
/ \tau _{GMC} ~\approx $ $3 \times 10^3$~(\lsol\ )~$E_{51} \tau _{4 \times
10^6}$ , where the dynamical time is taken to be a cloud crossing time,
$\tau _{GMC}$ = $R_{GMC}/ \sigma _{GMC}$ and $\sigma _{GMC}$ is the cloud
velocity dispersion. Since the above infrared transitions are important
coolants in the post--shock layers of these low velocity shocks, we expect
that future observations of clouds like Perseus will show extended bright
emission in these infrared transitions. 

Observations of young stellar populations show that stars form from GMCs
over a period of the order of 10 to 20 Myrs (Blaauw 1991) which implies that
clouds must survive for at least this long. Since this time--scale
exceeds the dissipation time, the internal motions must be regenerated. 
There are several possible sources for such energy generation. When
massive stars are present, their radiation, winds, and supernova
explosions inject large amounts of energy into the surrounding gas.
However this amount of energy is far in excess of what is required to
balance the dissipation of kinetic energy in shocks. Massive stars are
likely to be responsible for cloud disruption. 

The second candidate energy source is outflows from low mass stars which
form more uniformly throughout the cloud (Strom et al.\ 1989; Strom, 
Margulis \& Strom 1989). The Perseus
molecular cloud contains only intermediate to low mass stars, which during
their first $10^5$ years of life produced jets and outflows. All the cores
studied here are known to have multiple outflows. Over a dozen outflows in
NGC1333, 8 groups of Herbig--Haro objects in L1448, and 2 outflows in B5
have been discovered. Outflows can provide the mechanism to stir the
clouds and balance the dissipation of the random supersonic motions. 

However, the main source of energy for the small and intermediate scale
differential velocity field may simply be cascading of energy from larger 
scales, where supernova activity is a likely source of
galactic turbulence (\cite{Korpi+98a,Korpi+98b,Korpi+99a}).
The dynamical times of inertial scale motions are short compared
to those of the larger scale motions. In the inertial range,
there is a quasi-stead flow of energy from larger scales towards the
energy dissipation scales.  The numerical simulations represents 
only piece of the inertial range, and hence, on these scales, energy
input at the largest wavenumbers is to be expected.

\section{Conclusions}

Padoan, Jones \& Nordlund (1997) and Padoan \& Nordlund (1998) have shown
that statistics of infrared stellar extinction (Lada et al.\ 1994) and
OH Zeeman measurements (for example Crutcher et al.\ 1993; Crutcher et al.\ 
1996) are consistent with the properties of supersonic 
random flows. Padoan et al.\ (1998) have calculated 
values of CO line intensity ratios and line width ratios, using 
synthetic spectra, that are very close to the values measured 
by Falgarone et al.\ (1991) and by Falgarone \& Phillips (1996) 
in quiescent regions. In the present work, we have shown that 
idealized turbulent flows, without self--gravity, stellar radiation, 
stellar outflows, or any other effect of star formation, can
also provide a description of statistical properties of molecular 
clouds that are actively forming stars. We have also shown, using the
relation between integrated antenna temperature and equivalent
line width, that the motions in the Perseus complex must be 
super--Alfv\'{e}nic, with the exception of the core B1, where in 
fact a strong magnetic field has been detected. This is a further 
confirmation of the super-Alfv\'{e}nic molecular cloud model 
proposed by Padoan \& Nordlund (1998), where the formation of cores 
with relatively strong magnetic field is also predicted. 

It can be concluded therefore
that super--sonic and super--Alfv\'{e}nic randomly forced turbulence
correctly describes the structure and dynamics of molecular clouds,
even when they are apparently affected by star formation. 
This cannot be truly surprising, because, no matter what the
energy sources are, the motions in molecular clouds must be
highly turbulent, due to the very large Reynolds number,
and statistical properties of turbulence such as probability
density functions are universal, both in nature and in 
computer simulations.

\nocite{Arons+Max75,Aspin+94,Bachiller+87,Bally+96,Bally+96a,Bally+97}
\nocite{Bally+Lada83,Bally+87,Bally+89,Bally+91,Billawala+97}
\nocite{Crutcher+93,Crutcher+96,Elmegreen85,Elmegreen97imf}
\nocite{Falgarone92,Blaauw91,Falgarone+Phillips91,Falgarone+91,Falgarone+Puget86}
\nocite{Falgarone+Phillips96,Ferrini+83,Fleck88,Fleck96,Goodman+89,Heiles84,Henriksen+Turner84}
\nocite{Herbig+Jones83,Hodapp+Ladd95,Juvela97,Sandell+Knee98,Lada+96,Lada+Lada95}
\nocite{Lada+94,Lada+93,Langer+89,Larson81,Larson95,Lord+96,McKee+Zweibel95}
\nocite{Mestel65,Miesch+Bally94,Mouschovias76a,Mouschovias76b,Padoan+97ext}
\nocite{Padoan+97imf,Padoan+97cat,Padoan+Nordlund98MHD,Parker73,Sandell+94}
\nocite{Scalo87,Scalo90,Strittmatter66,Strom+74,Strom+89,Strom+89a,Whitworth79}
\nocite{Wright+91,Zweibel+Josafatsson83}
\nocite{Cernis90}

\acknowledgements

We thank Alyssa Goodman and the anonymous referee for their useful comments. 
Computing resources were provided by the Danish National Science
Research Council, and by the French `Centre National de Calcul
Parall\`{e}le en Science de la Terre'. PP is grateful to the Center for
Astrophysics and Space Astronomy (CASA) in Boulder (Colorado) for the warm
hospitality offered during the period in which this paper has been
written. JB and YB acknowledge support from NASA grant NAGW--4590 (Origins)
and NASA grant NAGW--3192 (LTSA). The work of MJ was supported by the
Academy of Finland Grant No. 1011055. 


\clearpage

\onecolumn

{\bf Figure captions:} \\

\begin{figure}
\centering
\leavevmode
\epsfxsize=1.0
\columnwidth
\epsfbox{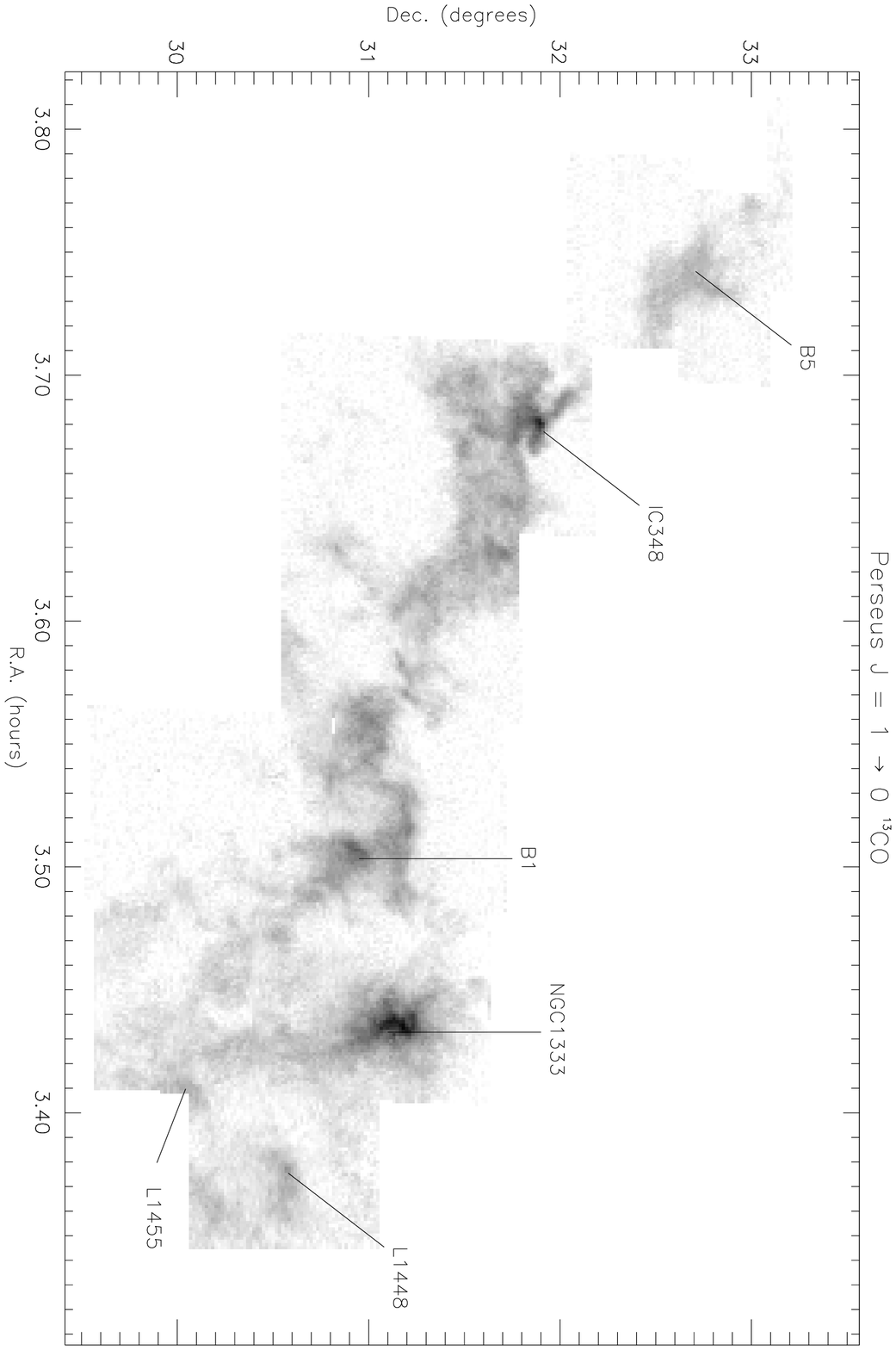}
\caption[]{}
\label{fig1}
\end{figure}

{\bf Figure \ref{fig1}:} Integrated antenna temperature of 
\Jtco\ for velocities 0 to 15~\kms\  \\

\begin{figure}
\centering
\leavevmode
\epsfxsize=1.0
\columnwidth
\epsfbox{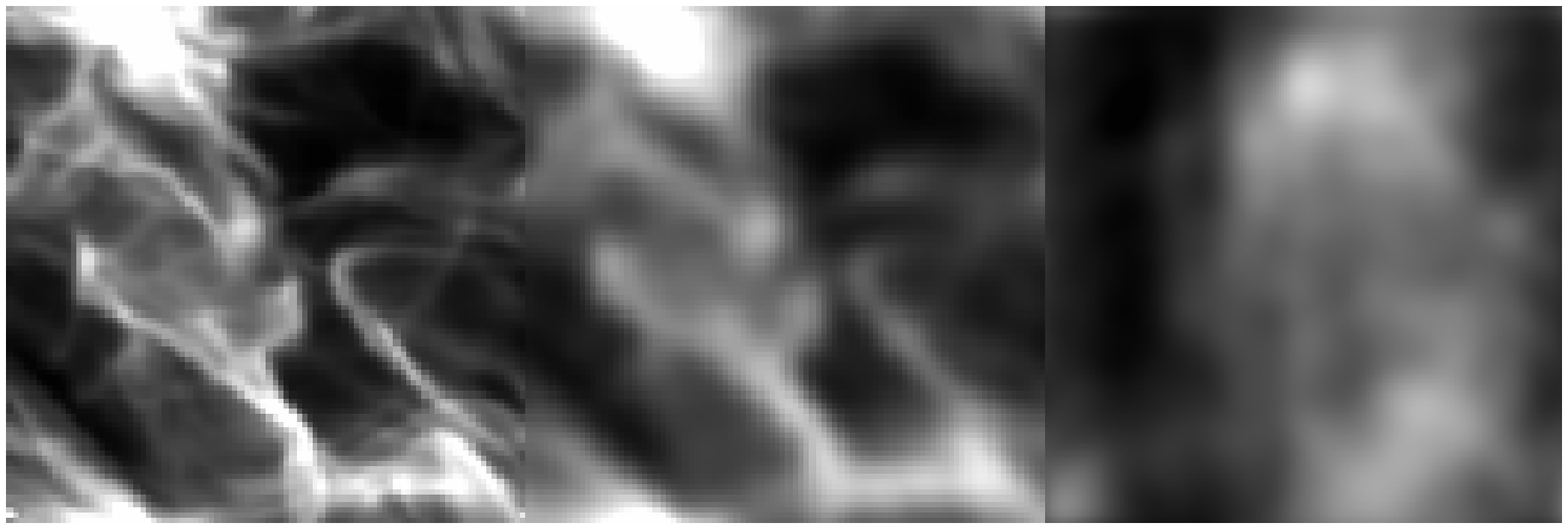}
\caption[]{}
\label{fig2}
\end{figure}

{\bf Figure \ref{fig2}:} Integrated antenna temperature of the model cloud 
(left and middle panels) and of L1448 (right panel). The middle and the 
right panels have been smoothed with a gaussian filter of width comparable
with the 100" observational beam. The left panel is also smoothed with a 
gaussian filter, but the width is comparable with the grid size. \\

\begin{figure}
\centering
\leavevmode
\epsfxsize=1.0
\columnwidth
\epsfbox{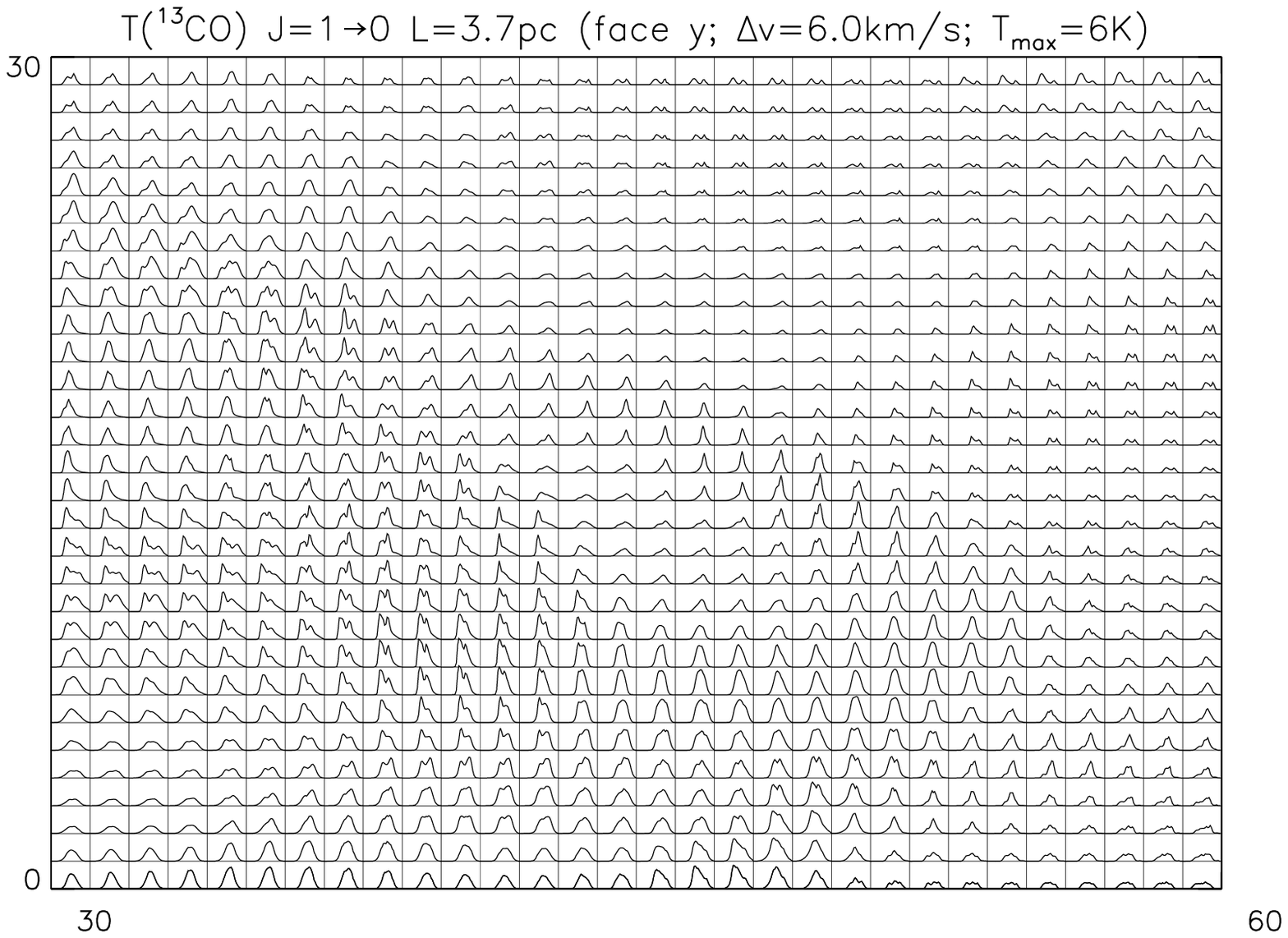}
\caption[]{}
\label{fig3}
\end{figure}

{\bf Figure \ref{fig3}:} A 30 $\times$ 30 map of individual \Jtco\ 
synthetic spectra from the model cloud. The velocity interval in the plots 
is 6.0~\kms\ and the antenna temperature ranges from 0 to 6 K. The original 
map contains 90 $\times$ 90 synthetic spectra. \\

\begin{figure}
\centering
\leavevmode
\epsfxsize=1.0
\columnwidth
\epsfbox{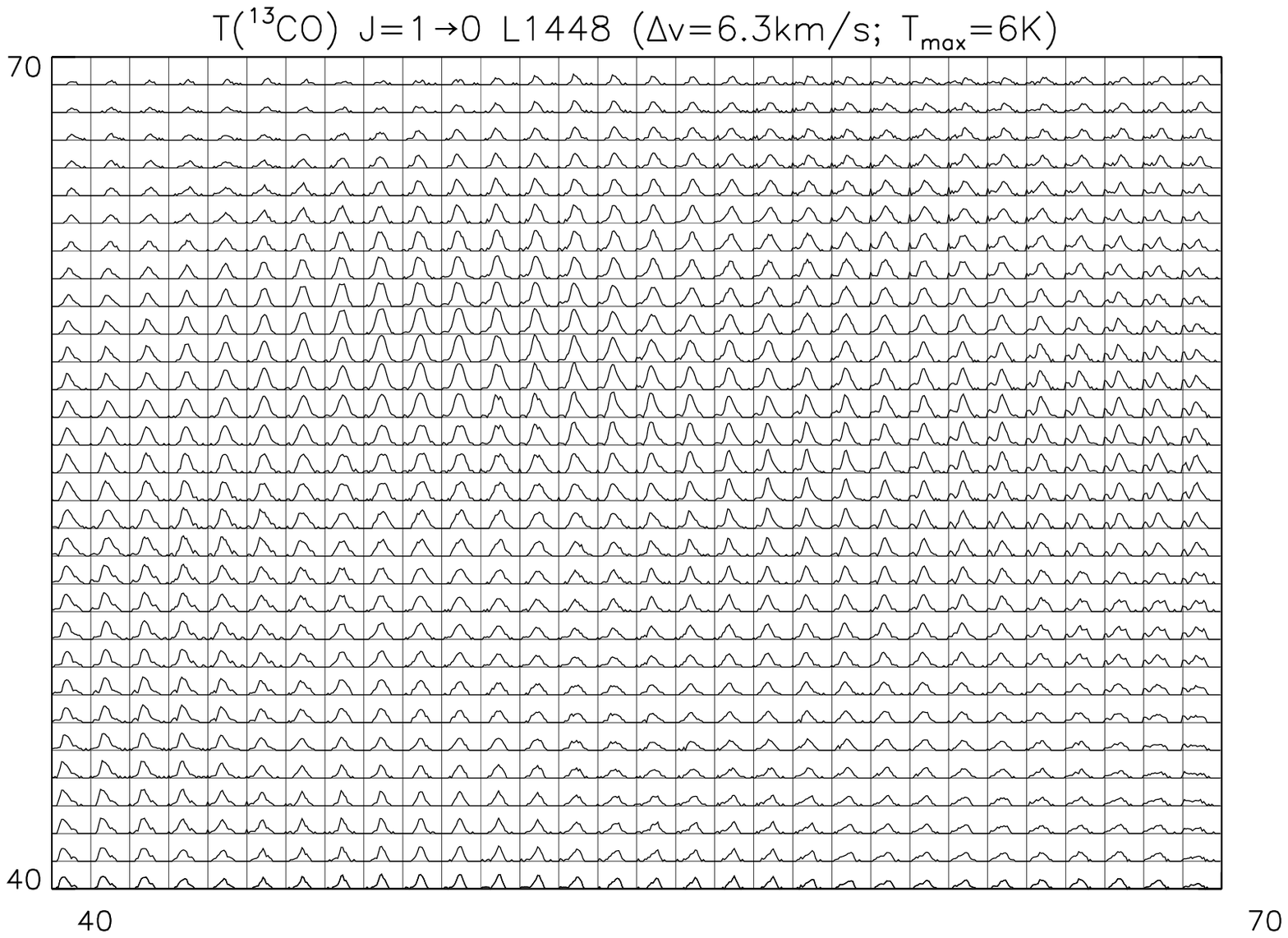}
\caption[]{}
\label{fig4}
\end{figure}

{\bf Figure \ref{fig4}:} A 30 $\times$ 30 map of individual \Jtco\ 
observed spectra of L1448. The velocity interval in the plots is 6.3~\kms\ 
and the antenna temperature ranges from 0 to 6 K.  \\

\begin{figure}
\centerline{\epsfxsize=13cm \epsfbox{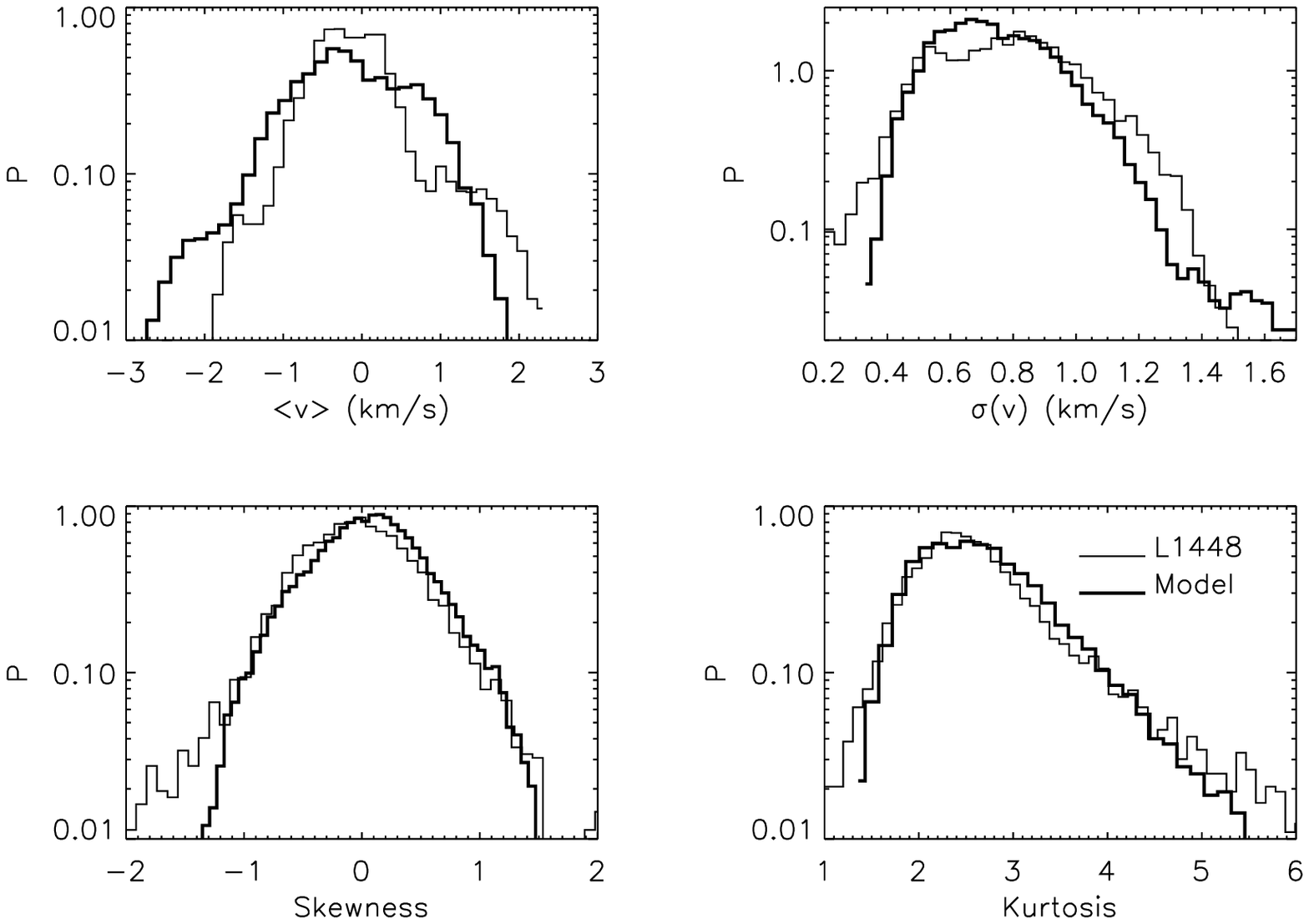}}
\centerline{\epsfxsize=13cm \epsfbox{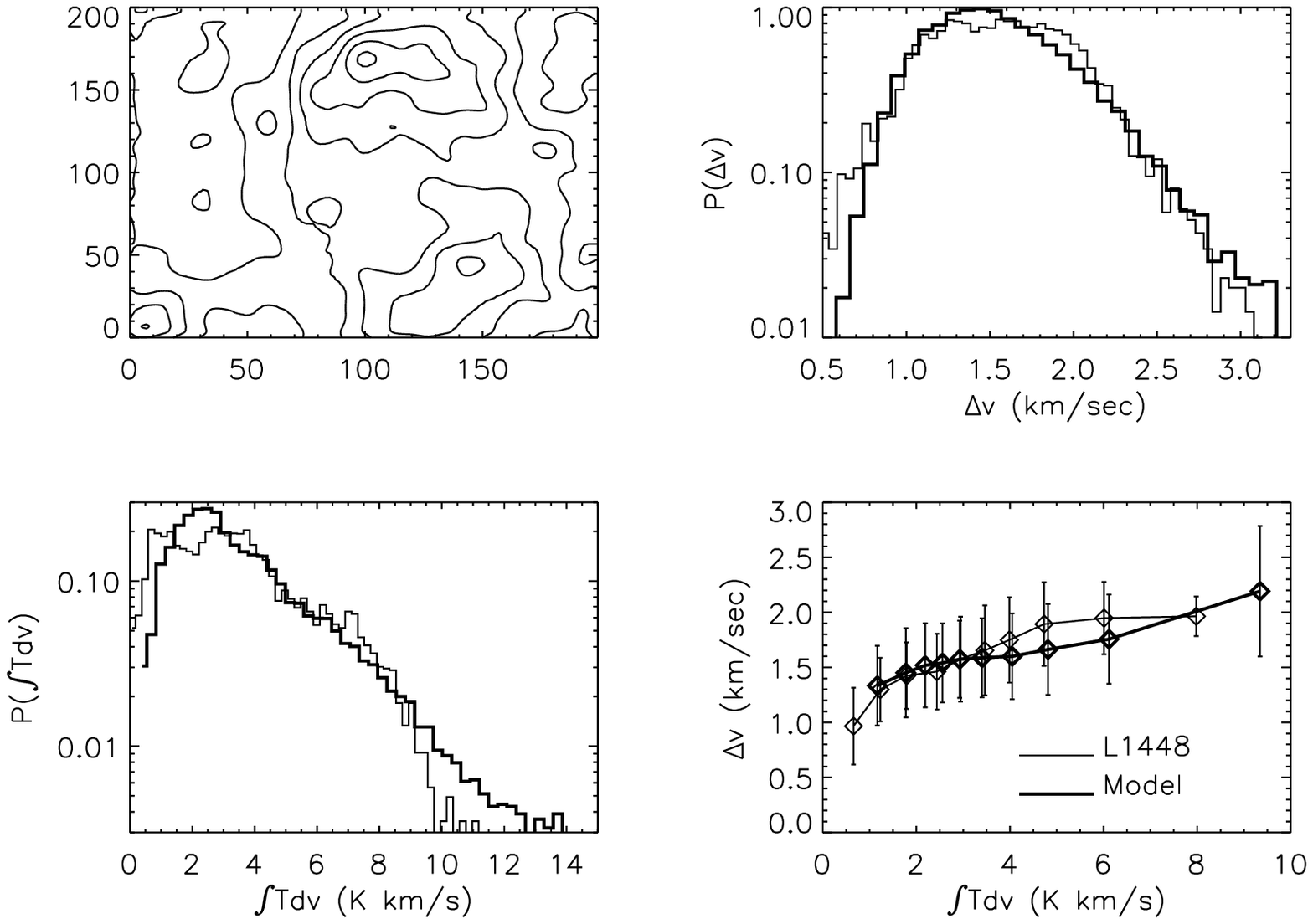}}
\caption[]{}
\label{fig5}
\end{figure}

{\bf Figure \ref{fig5}:} The top four panels: 
Distribution of centroid velocities (upper left);
distribution of line widths (upper right);
distribution of skewness (lower left); and
distribution of kurtosis (lower right). 
The bottom four panels:
Contour map of integrated temperature of the selected region (upper left);
distribution of equivalent width (upper right); 
distribution of integrated antenna temperature (lower left);
and equivalent width versus integrated antenna temperature (lower right).
Thick lines are computed using spectra of single lines of sight 
from the model cloud. Thin lines are from the region around the core L1448. \\

\begin{figure}
\centerline{\epsfxsize=13cm \epsfbox{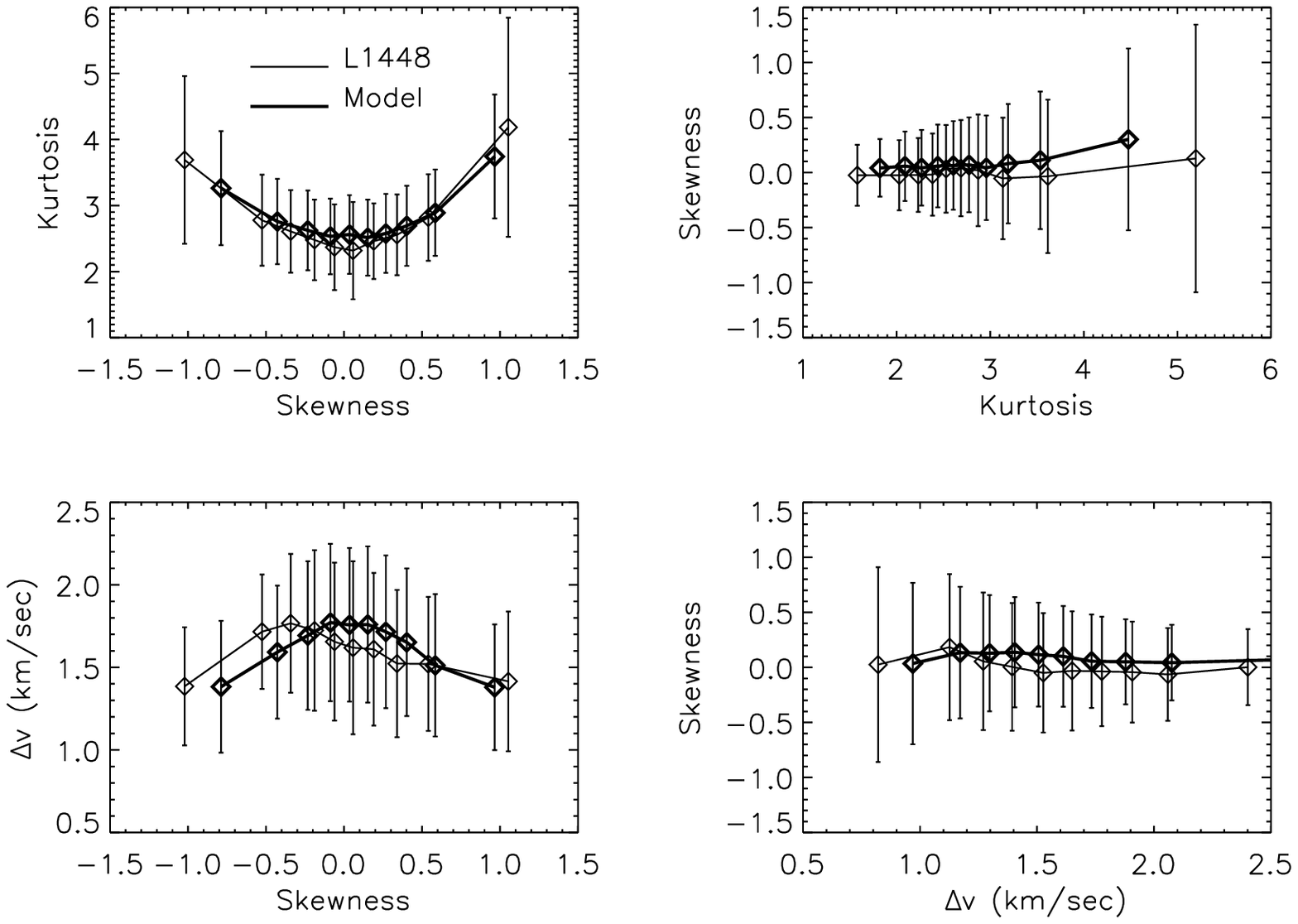}}
\centerline{\epsfxsize=13cm \epsfbox{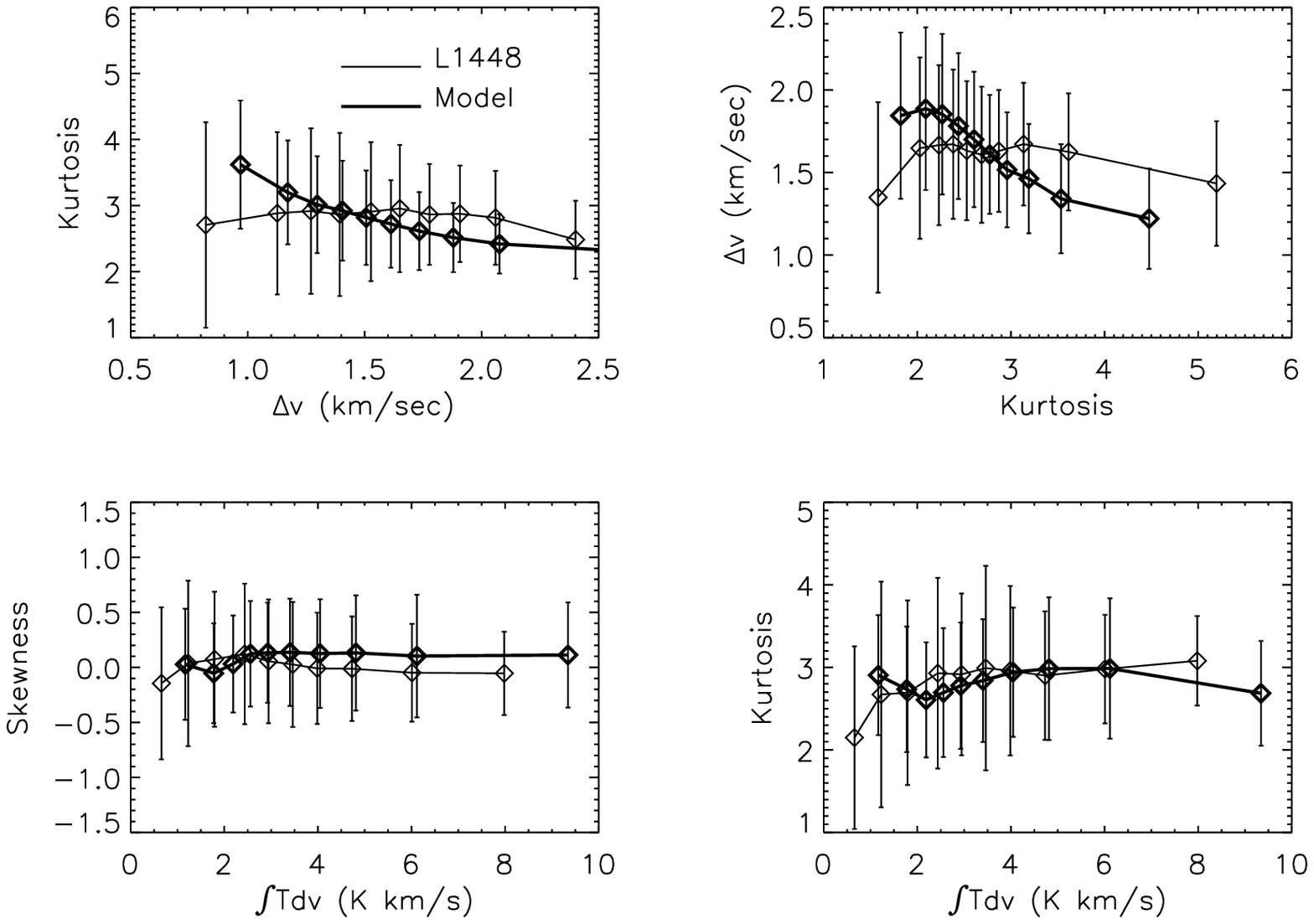}}
\caption[]{}
\label{fig11}
\end{figure}

{\bf Figure \ref{fig11}:} Correlations between different properties of the
theoretical spectra (thick lines) and of the spectra of the region around 
the core L1448 (thin lines). \\

\begin{figure}
\centerline{\epsfxsize=13cm \epsfbox{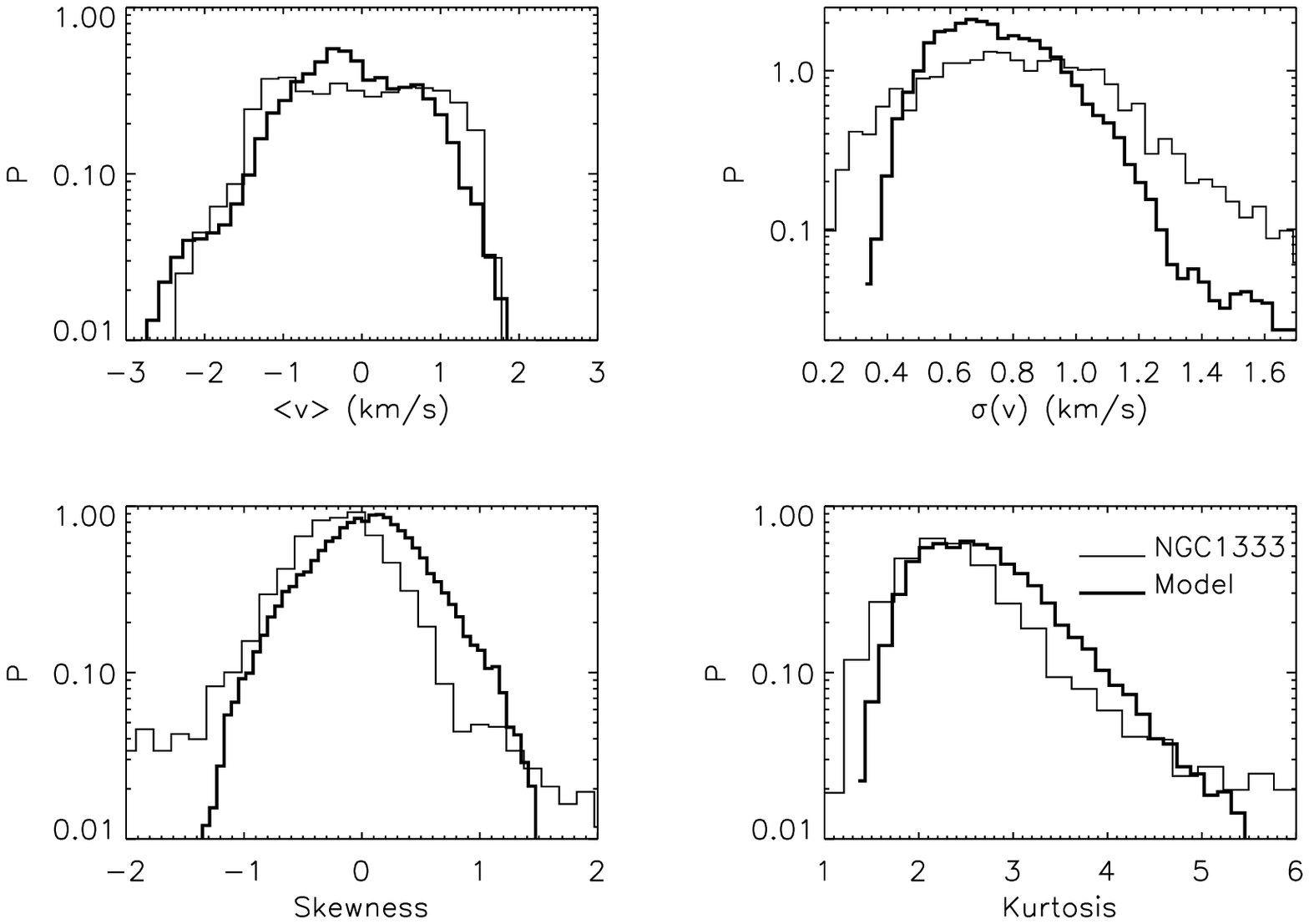}}
\centerline{\epsfxsize=13cm \epsfbox{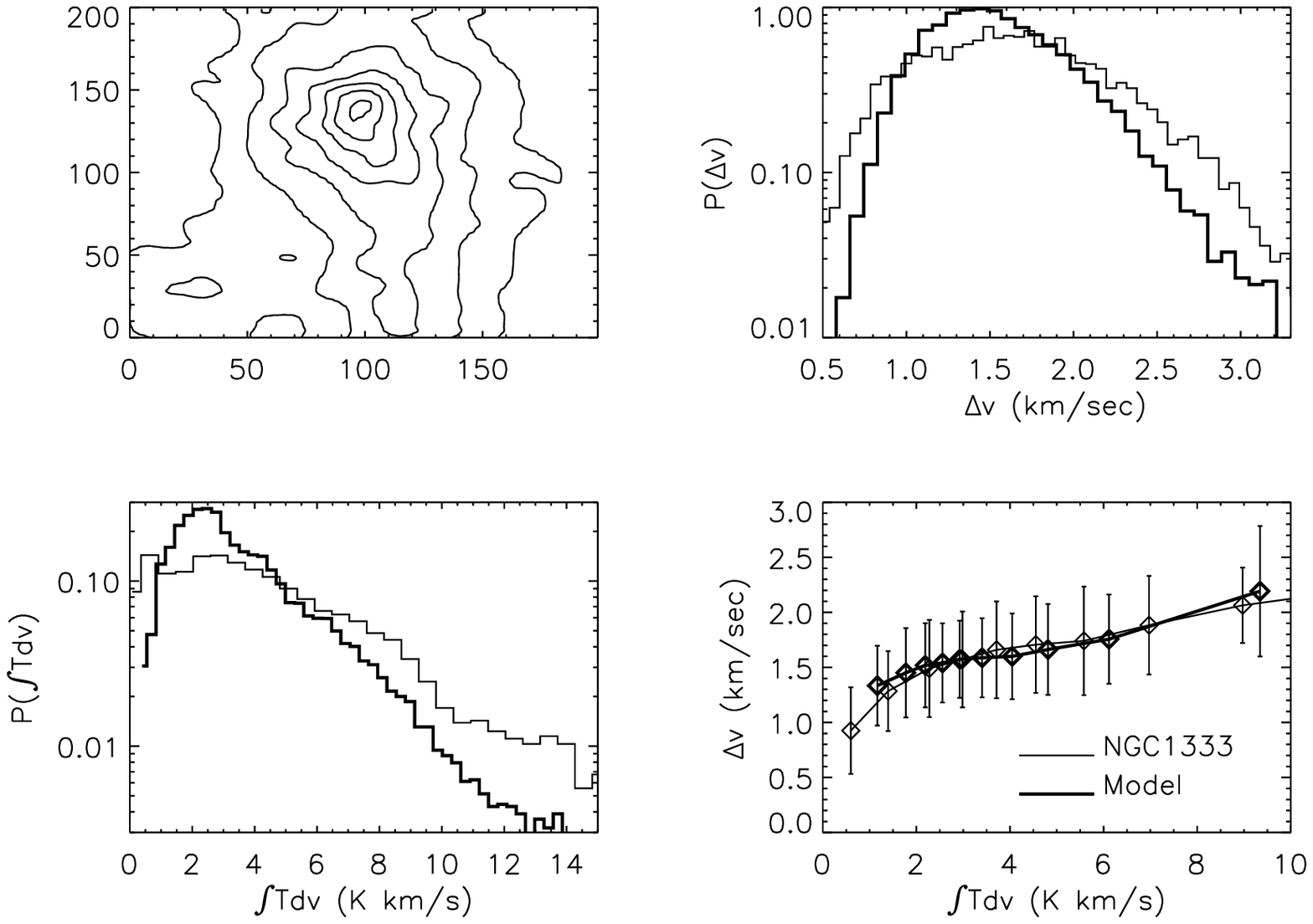}}
\caption[]{}
\label{fig6}
\end{figure}

{\bf Figure \ref{fig6}:} Same as Fig.~\ref{fig5}, but for the region around 
the core NGC1333. \\

\begin{figure}
\centerline{\epsfxsize=13cm \epsfbox{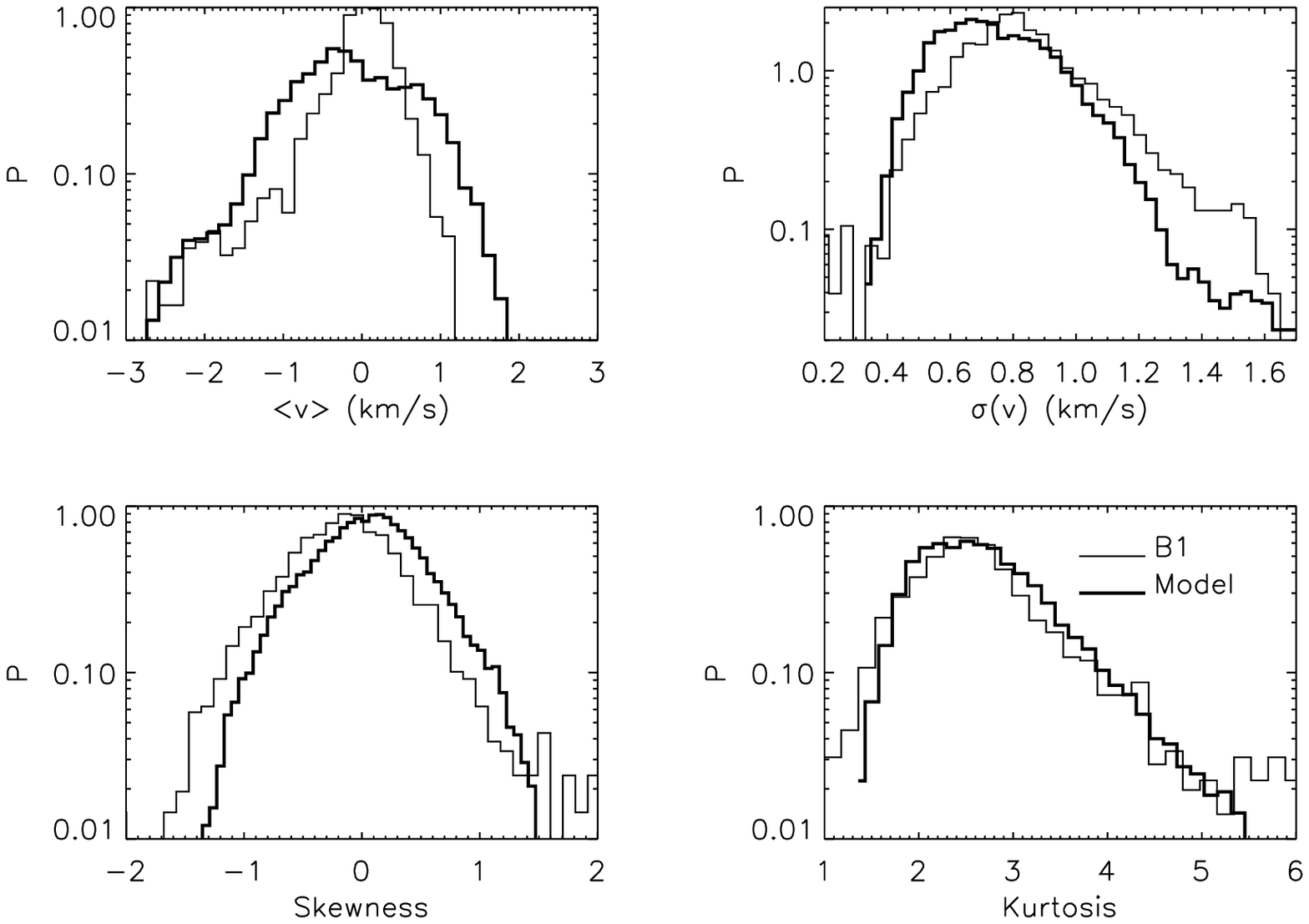}}
\centerline{\epsfxsize=13cm \epsfbox{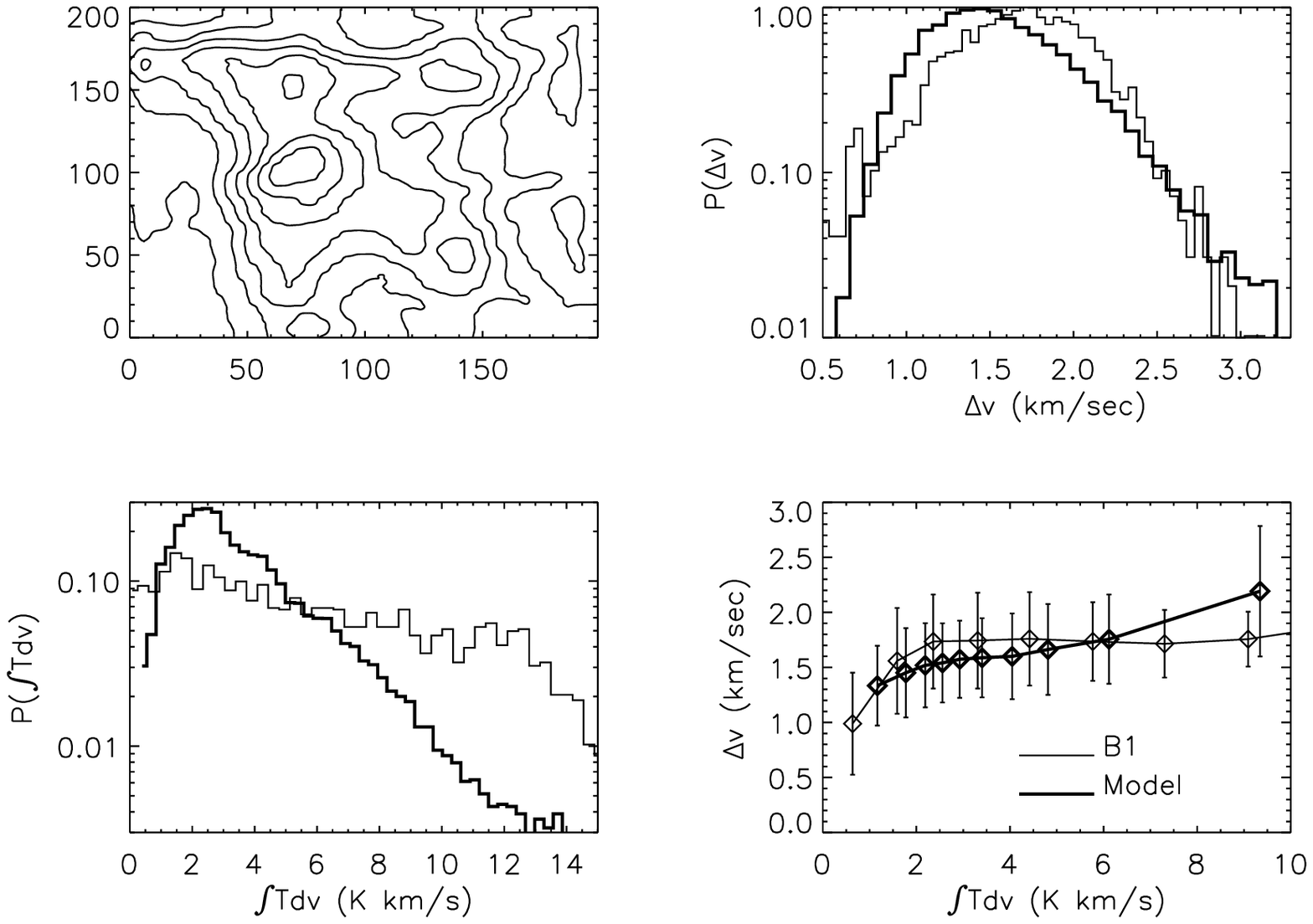}}
\caption[]{}
\label{fig7}
\end{figure}

{\bf Figure \ref{fig7}:} Same as Fig.~\ref{fig5}, but for the region around 
the core B1. \\

\clearpage

\clearpage

\clearpage

\clearpage

\clearpage

\clearpage

\clearpage

\clearpage

\end{document}